# Building Dynamic Ontological Models for Place using Social Media Data from Twitter and Sina Weibo

Ming-Hsiang Tsou [ab]*, Qingyun Zhang [ab], Jian Xu[ab], Atsushi Nara [ab], Mark Gawron [ac]

[a]*The Center for Human Dynamics in the Mobile Age, San Diego State University, U.S.A.*

[b]*The Department of Geography, San Diego State University, U.S.A.*

[c]*The Department of Linguistics, San Diego State University, U.S.A.*

mtsou@mail.sdsu.edu, rick8600@gmail.com, jxu9383@sdsu.edu, anara@mail.sdsu.edu, gawron@mail.sdsu.edu.

*: contact author.

# Abstract

Place holds human thoughts and experiences. Space is defined with geometric measurement and coordinate systems. Social media served as the connection between place and space. In this study, we use social media data (Twitter, Weibo) to build a dynamic ontological model in two separate areas: Beijing, China and San Diego, the U.S.A. Three spatial analytics methods are utilized to generate the place name ontology: 1) Kernel Density Estimation (KDE); 2) Dynamic Method Density-based spatial clustering of applications with noise (DBSCAN); 3) hierarchal clustering. We identified feature types of place name ontologies from geotagged social media data and classified them by comparing their default search radius of KDE of geo-tagged points. By tracing the seasonal changes of highly dynamic non-administrative places, seasonal variation patterns were observed, which illustrates the dynamic changes in place ontology caused by the change in human activities and conversation over time and space. We also investigate the semantic meaning of each place name by examining Pointwise Mutual Information (PMI) scores and word clouds. The major contribution of this research is to link and analyze the associations between place, space, and their attributes in the field of geography. Researchers can use crowd-sourced data to study the ontology of places rather than relying on traditional gazetteers. The dynamic ontology in this research can provide bright insight into urban planning and re-zoning and other related industries.

**Keywords:** *Social Media, Ontology, GIS., Spatiotemporal Analysis, Semantic Analysis, Visualization.*

# 1. Introduction

Place and space are two important concepts in geography. They are distinct but interior interrelated. Place subjectively carries human thoughts and experiences. There are many definitions for a place. People can define a place by using the administrative boundaries of cities, counties, or states, such as La Mesa, California, or the U.S.A. A place can also be defined or referred by gazetteers or points of interest (P.O.I.s), such as San Diego State University or Sea World. Some places' names have been used for a long time in historical documents, while some places' names are created recently and are relatively new, such as a new restaurant opening in San Diego downtown. Different from the definition of space, people define places with some similar cognitive boundaries or personal experiences in their minds.

Social media are deeply ingrained in human lives and experiences. Most social media users write their messages on their cell phones or mobile devices. Some users may turn on the geolocation information when they post their social media messages, which can be labeled as "geo-tagged" messages and be analyzed by researchers. Social media services, such as Twitter or Instagram, provide application programming interfaces (APIs) that allow developers to collect a portion of social media data randomly or systematically from their servers. Cloud computing and high-performance servers allow researchers to store and analyze the large size of social media datasets.

This study collected and analyzed geo-tagged social media data containing same place. Geo-tagged social media data can be used to analyze the connections and relationships between place and space because these data can record both human thoughts regarding places and activity activities in space through geotags. These data that contain identical place were stored in a database management system to build a dynamic ontology of place. This study utilized the "bottom-up" approach (Keßler et al., 2009) and crowd-sourced data (social media) to study the ontology of places rather than relying on traditional gazetteers.

**Direction and Objective**

This research focuses on portraying ontological boundaries of place by using geo-tagged social media data from multiple sources as well as observing the temporal changes of the boundaries associated with place. Dynamic ontological boundaries can be defined by crowd-sourced geo-tagged social media data, which can illustrate the dynamic change of place from a spatiotemporal perspective. Unlike the traditional definition of place from gazetteers or data dictionaries, this new dynamic ontology framework can provide human-centered definitions of place and dynamic updates of place ontology.

Specifically, we are trying to:

1. use social media data instead of traditional data (e.g., google map) to detect place feature characteristics, and explored the significance and value of social media in place feature detection
2. used different social media (Twitter and Weibo) data for feature detection, temporal-spatial analysis, and extracted word cloud information for each place name studied to analyze the features of the place and compare the differences and associations.

# 2. Literature Review

Place can often be defined as a location to which meaning is assigned through human experience, which means it has multidisciplinary scientific significance. By integrating references to human activities within urban spaces, we can observe the emergence of distinctive themes characterizing different places, and thus identify places by their discernible sociocultural features. The definition of place keeps evolving, and over time, places are given new meanings, such as reflecting urban dynamics, evolving social culture, perceptions, or major events. In Cranshaw et al. (2012) study, the "character" of urban areas did not depend only on the type of place, but also on the people who choose to make the area a part of their daily life, which also reflects the dynamic activity patterns of the place. Some researchers have used a combination of statistical methods to extract spatio-temporal and semantic features from Twitter. Hu et al. (2015) used Dynamic Method Density-based spatial clustering of applications with noise (DBSCAN) clustering algorithm to identify urban area of interest (AOI) by extracting textual tags and preferable photos from social media (Flickr). For example, Jenkins et al. (2016) used Twitter data to extract two clusters of entertainment and recreation in Manhattan, New York, and the corresponding physical buildings. Hu et al. (2015) analyzed the behavioral characteristics of people in AOIs by using geotagged images on Flickr to understand and summarize the characteristics of AOIs, thus exploring the dynamics of AOIs, and identifying potentially similar AOIs.

Tuan (1979) introduced some fundamental ideas about place and space. He mentioned that natural geography is defined by place and space. On the one hand, the meaning of place was more productive and divergent than the meaning of space. Space was objective and sensed-bound to the facts of the physical realm. On the other hand, place could be considered as a single unit that is connected by a circulation net. Place also embodied humans' experiences and thoughts, which embrace many kinds of "personalities" and "spirits." These can only be sensed by human beings. Janowicz and Kebler (2008) identified a gazetteer as a dictionary of places that included three components: the places' names, the footprint of the places, and the features of the places. Each component can refer to either of the two other components. Gruber (1995) defined ontology as an explicit specification of a conceptualization. He mainly described the design of ontologies.

Gao et al. (2017) identified continuous boundaries and constructed polygon representation of cognitive regions using a data-synthesis-driven approach. They identified the cognitive regions of NoCal (Northern California) and SoCal (Southern California) based on social media data from multiple platforms (Twitter, Flicker, and Instagram). Li et al. (2012) constructed places from spatial footprints by using the interaction between photographs and geographic landscapes. The dictionary of a place can include multiple names, footprints, features, and can provide components of complex reasoning services, such as the identity assumption service for historic places.

One place can be either independent or connected to other established places. Li et al. (2012) put forward that the traditional official gazetteer had some limitations. First, it has a very narrow target audience, such as authorized persons such as surveyors and cartographers. Second, the maintenance of traditional gazetteer is expensive and time-consuming. By contrast, both social media and volunteered geographic information (V.G.I.) can provide more flexible and dynamic knowledge of geospatial information (Hu et al. 2015). The place ontology, which includes place and spatial footprints and can identify geographical awareness (Han et al. 2015, Scheider and Janowicz, 2014), is ingrained by human perception and deserves to be further explored using new forms of data.

Social media platforms can provide up-to-date, concise, and insightful big data. Big data is characterized as high volume, unstructured, high-velocity, wide variety, complex dataset. Big data can be processed and analyzed computationally and cost-effectively to retrieve useful information, such as patterns and trends, indicating human behaviors or activities. Social media, as one of the primary sources of big data, can reflect the dynamic thoughts and conversations in our society. Therefore, a place name ontology created by social media will become dynamic and can change over time. People post their statuses with geospatial information by using their mobile devices(Shaw et al. 2013). Tsou et al. (2013) noted that social media data created by the general public could be used to track their digital footprints.

From a human perspective, social media can be used to observe human activities since social media data record human interaction. Tsou et al. (2015) discussed that social media data are a visible indicator of the diffusion of innovations or information. Social media data can be used to monitor an outbreak event spatially. Jiang et al. (2015) indicated that many researchers utilize social media to study disease outbreaks, natural disasters, and sports events, such as the Olympic Games and the World Cup. Tsou et al. (2015) developed a social media analysis platform, called "SMART dashboard" to monitor disease outbreaks and other events. The platform can monitor analysis on many topics such as flu outbreaks, wildfires, and social movements. Tsou et al. (2015) used social media data (Twitter data) to compare the weekly result of FluView provided by the Centers for Disease Control and Prevention (C.D.C.) and found that the weekly pattern from flu-related tweets is highly correlated with the weekly pattern from the C.D.C. FluView. Jiang et al. also found a strong correlation between two data sources - the Chinese social media data (Sina Weibo) and the AQI data in Beijing gathered from the China Ministry of Environmental Protection about air pollution conditions. After data filtering, the consistency of verifiable data is quite high in most scenarios. Earle et al. (2012) highlighted that social media gives a very intuitive natural disaster awareness. For example, when an earthquake comes, Twitter feeds rapid updates, sometimes even faster than seismic stations. Consequently, social media data is ideal for the ontological analysis of place. There are many ways of collecting social media data. In general, the required social media data can be queried by keywords via application programming interfaces (API).  API is an integrated online service for collecting raw social media data. Data filtering and cleaning procedures are also crucial to improve the accuracy of social media analysis. A data cleaning algorithm needs to be developed with machine learning methods (Jiang et al. 2015 & Tsou 2015) to filter raw social media data. Twitter and Sina Weibo provide robust data collection platforms that indicate different user behaviors and activities, which are further reflected in divergent ontological analysis results.

Cartography and geo-visualization are techniques that represent data visually and spatially. Visualization is widely applicable in many geographic research activities (MacEachren & Kraak 2001). With an effective visualization method, it can facilitate effectively analyze and explanation of data. Gao et al. (2017) discussed two algorithms for boundary portraying that are effective in their project. The Standard Deviation Ellipse (S.D.E.) and Fuzzy-Set-Based (F.S.B.) classification and identification methods were applied to portray the place of Santa Barbara Court and Harvard University and then used to figure out the best option. Elbatta and Ashour (2013) declared that density-based methods such as Density-based spatial clustering of applications with noise (DBSCAN) have the advantages of detecting arbitrary clusters forms and automatically filtering noises that are created by systematic errors. This research utilized the Kernel Density method, DBSCAN, and hierarchal clustering to build the boundaries of place.

Space and time are fundamental elements for observing events that represent the ontological framework. When an event happens, people are concerned about the time and location. Publications of space-time analysis have increased significantly from 1949 to 2013. An et al. (2015) followed Einstein's notion that an event cannot happen everywhere and anytime. Not only are geographers interested in this type of research, but it is popular in many other disciplines as well. It can be applied to many sub-disciplines of geography. A space-time statistical model can explain the variations, represent reality, predict the future trend, and be leveraged in decision making. For example, Huang and Wong (2015) use the S.T.P. (space-time path) to portray the human movement by utilizing G.I.S., which demonstrates individual periodic behaviors. In the HDMA center, people have been interested in studying the trajectory of human dynamics and disease outbreak trends to make predictions and further analysis. The dynamic of the boundaries can intuitively show that place ontology is not static, but rather follow seasonal patterns.

Overall, the previous studies illustrate the feasibility of using social media data to perform a spatial-temporal analysis of place name ontology. There are essential semantic meanings associated with each place name, which can be retrieved from crowd-sourced social media data. This research studied the relationship between different components of place ontology: place, locations and boundaries, and associated attributes.

# 3. Data collection and Methods

There are two case studies for this research. We used Twitter data to build the place name ontology for nine place in San Diego County, California, United States in the first case study ("San Diego". "university", "El Cajon", "La Mesa", "Chula Vista", "SDSU", "Sea World", "I-5", "I-8"). In the second case study, Weibo data were used to create eight different place name ontologies in Beijing, China ("Chaoyang" (朝阳), "Wangfujing" (王府井), "Haidian" (海淀), "Peking University" (北京大学), "Tiananmen square" (天安门广场), "Chang'an avenue" (长安街), "The 3rd ring road" (三环), "university" (大学)).

Data from the two social media platforms (Weibo and Twitter) were collected through their public application programming interfaces (APIs). Only geo-tagged social media data were collected for building dynamic ontology of place in San Diego and Beijing. This research utilized three main spatial analysis methods in order to analyze the dynamic ontological representations of place: 1). Kernel Density Estimation (KDE); 2) Density-based spatial clustering of applications with noise (DBSCAN); and 3) hierarchal clustering.  Figure 1 illustrates the main procedures in this research.

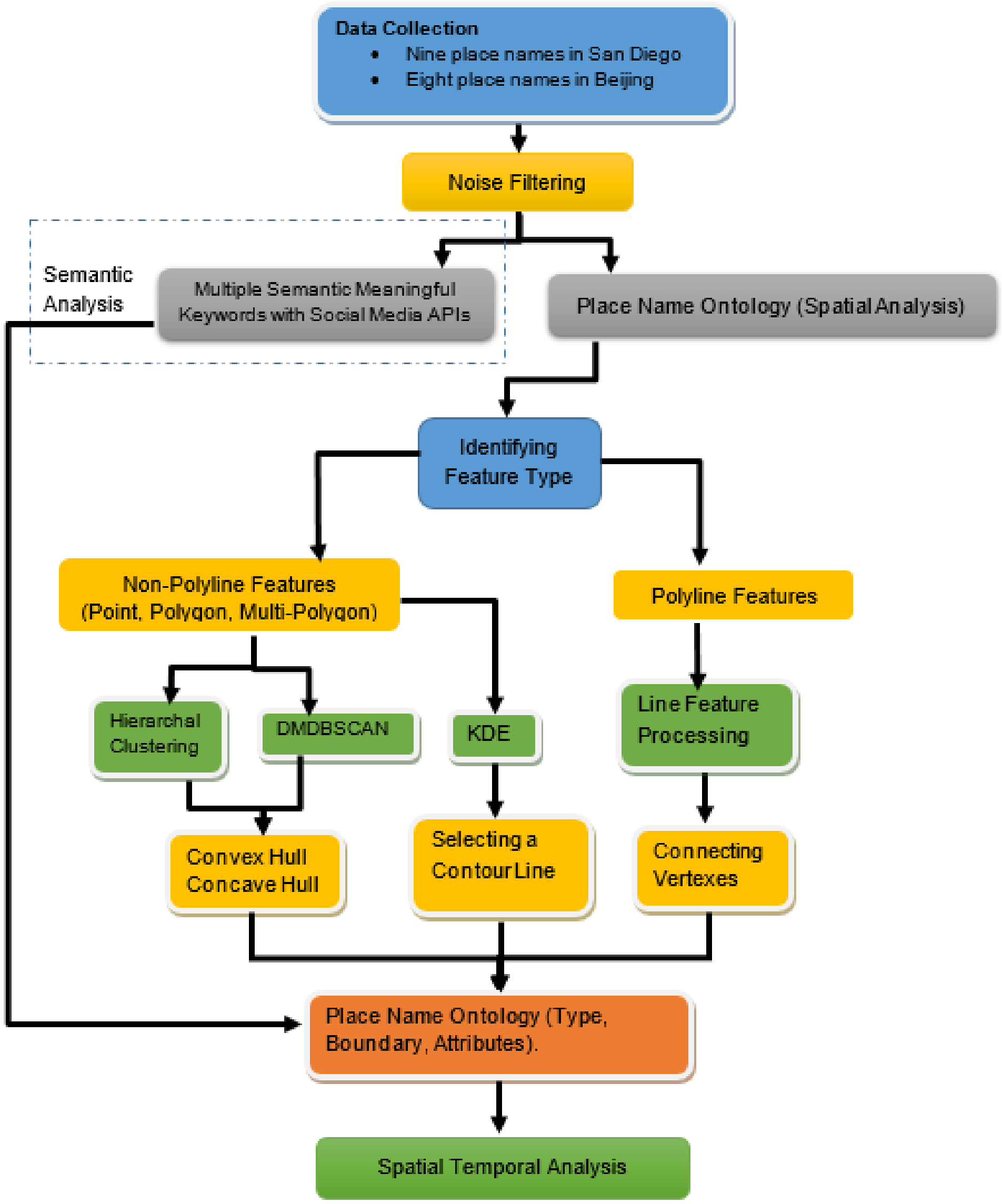

***Figure 1***. *The dynamic ontological analysis procedures for different place*

## 3.1 Data Collection

As shown in Figure 1, the dynamic ontology model requires geo-tagged social media data. Twitter streaming API collects instant Twitter data randomly, continuously and systematically, and stores the data on MongoDB at our server in JSON format. We collected two years of geo-tagged tweets from January 2015 through December 2016 within the boundary of San Diego County. Sina Weibo Open APIs have been used to collect the place associated with Weibo messages for eight place in Beijing, China. The collected geo-tagged Weibo messages are stored on MongoDB on Wuhan University's local server. The database contains geo-tagged Sina Weibo data from October 2013 through September 2014.

***Table 1.*** *The place chosen for building the dynamic ontology*

| Social Media Platform | City District | Municipal Public Places | Line Features | Place Name Hierarchy |
|---|---|---|---|---|
| Twitter | El Cajon, La Mesa, Chula Vista | SDSU, Sea World | I-5, I-8 | University, San Diego |
| Weibo | Chaoyang (朝阳), Haidian (海淀) | Peking University (北京大学), Tiananmen square (天安门广场), Wangfujin (王府井) | Chang'an avenue (长安街), The 3rd ring road (三环) | University (大学) |

The administrative boundary shapefile layers have been downloaded from the SANDAG website that records many established cities' boundaries. The rest of the administrative boundaries are digitized based on the basemaps (Streets or Tianditu map). The administrative boundaries serve as references.

Noise filtering is a critical component of the analysis task because many Weibo posts are posted by robots. The filtering process has two stages: First, it identifies possible bots' user accounts. Second, it removes those bots' posts. There are two major sources of noise. The first source is system error when collecting the data. These tweets are usually located outside of the original defined bounding boxes. The second source of noise is commercial bot and cyborg tweets. These tweets are most often advertisements. Each tweet record has metadata that indicates the source of the tweet. Tweets created by bots can be easily identified. According to previous research from Tsou (2017), 29.42% of collected tweets in San Diego have been recognized as noise data. Since noise in the data can significantly impact the result in this research, noise filtering has been processed on Twitter data based on tweets' sources. We removed those tweets outside the research boundaries and advertising tweets. We preprocessed Weibo data and only used the original posts to avoid noise. None of the geo-tagged Weibo posts contain advertisements or reposts.

***Table 2***. *The Data Noise Information*

| Social Media Platform | Original Number of Microblog Posts | Noise Posts Number | Final Posts Number | Noise Percentage (%) |
|---|---|---|---|---|
| Twitter | 7,619,307 | 864,477 | 6,754,830 | 11.34 |
| Sina Weibo | 11,951,385 | 0 | 11,951,385 | 0 |

## 3.2 Feature Type Identification

KDE is an efficient and popular algorithm for calculating the polygon features' densities. We normalized the KDE raster because certain regions have high tweet density due to high population density. In this study, the raster KDE of all geo-tagged tweets (tweets' population densities) in the database is the ground raster for normalization. For example, there are relatively low tweet population densities in some regions, while the tweet densities under specific place name keywords are relatively high. The differential value between the ratio of tweet density over the maximum kernel value of the keywords and the ratio of tweet population density over the maximum kernel value of tweet population density is crucial for generating hotspots on the map. Map algebra is utilized for the normalization. ArcMap has built-in "Map algebra" tool. The formula for KDE normalization is provided below. Formula 2 shows the differences or changes between two different raster layers.

$$Differential\,Value\ = \frac{KeywordA}{MaxKernelValueofKeywrodA} - \frac{KeywordB}{MaxKernelValueofKeywrodB}(1)$$

(Tsou 2013)

*Where "KeywordA" stands for the cell values of the keyword A's KDE raster layer. The formula normalizes the value by dividing the maximum cell value of the KDE raster layer. If certain regions' differential values are higher than 0, then the keyword A is dominated in these regions rather than keyword B and vice versa.*

As shown in figure2, first, this study used the default value as a search radius and decided to use 100 cell size by default on the KDE tool. If the points of the dataset are densely distributed, the default search radius can be relatively smaller, and the spatial outliers will be isolated. Second, if the bandwidth is large, this place name should be categorized as a line feature. By contrast, using tweets of normal polygon feature type places, such as El Cajon and SDSU, can generate the KDE raster that has a normal search radius. To find out the default search radius for each place, the execution logs were examined while performing a kernel density each time. Table 3 showed each type of feature in Beijing and San Diego, as well as their search radius threshold.

There is a general rule for detecting polyline and non-polyline features in two different case studies:

- In San Diego, if the default radius is larger than 10,000 meters, it will be considered a polyline feature.

- In Beijing, if the default radius is larger than 1,000 meters, it will be considered a polyline feature since Weibo is denser than tweets in San Diego

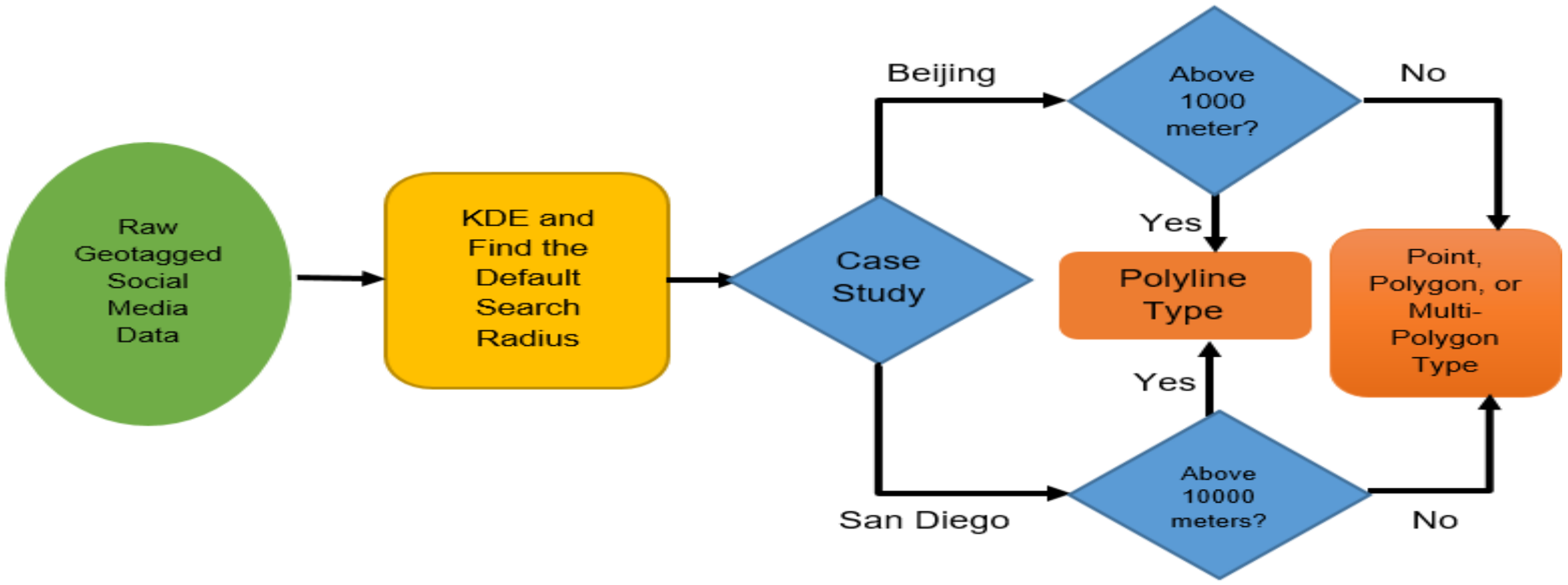

***Figure 2**. The Flow Chart of Feature Type Detection*

***Table 3**. The default search radius of KDE for place in Beijing and San Diego*

| Original Feature Type | Case Study | Placename | Default Search Radius (m) | Detected Feature Category |
|---|---|---|---|---|
| Point | San Diego | SDSU | 1055.2637 | Non-Polyline |
| | | Seaworld | 2855.033 | Non-Polyline |
| | Beijing | Wangfujin | 233.42 | Non-Polyline |
| | | Peking University | 692.76 | Non-Polyline |
| | | Tian'anmen Square | 269.5756 | Non-Polyline |
| Polyline 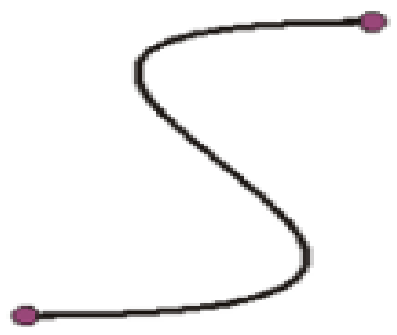 | San Diego | I-5 | 33525.77 | Polyline |
| | | I-8 | 11641.245 | Polyline |
| | Beijing | 3rd Ring Rd | 1424.168 | Polyline |
| | | Chang'an Ave | 242.1619 | Non-Polyline |
| Polygon 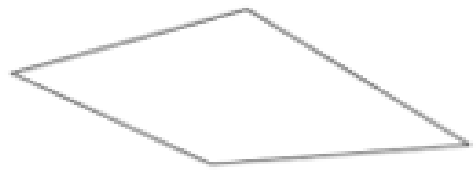 | San Diego | Chula Vista | 675.76 | Non-Polyline |
| | | La Mesa | 292.54 | Non-Polyline |
| | | El Cajon | 479.13 | Non-Polyline |
| | | San Diego | 552.0748 | Non-Polyline |
| | Beijing | Chaoyang | 673.46 | Non-Polyline |
| | | Haidian | 692.76 | Non-Polyline |
| Multiple Polygons 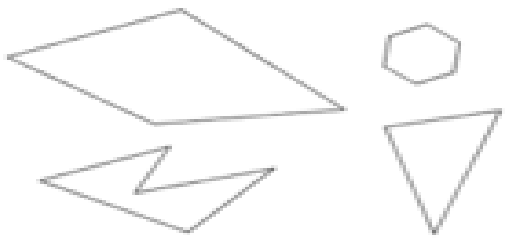 | San Diego | University | 7077.94 | Non-Polyline |
| | Beijing | University | 885.48 | Non-Polyline |

In this research, we also introduce Hierarchal Clustering and Dynamic Method Density Based Spatial Clustering of Applications with Noise (DMDBSCAN) to analyze the spatial footprints of place. Hierarchal clustering is a clustering strategy that minimizes within-group variance and maximizes group variance to get the appropriate clusters. DMDBSCAN is a quick and smart way to cluster existing points. The tweet points that are selected for creating the largest cluster are used to form the boundary of the polygon feature place name. The algorithm of the model requires the input of the total amount of geo-tagged tweet points and their standard distance.

## 3.3 Spatial-Temporal Analysis

In the spatial-temporal analysis of this study, the monthly variations have been examined. The monthly data from June 1, 2015, to December 2015 was collected. In Figure 3a and b, we can identify that the ontological area changes in three established cities (Chula Vista, El Cajon, and La Mesa) and two other places (S.D.S. U and Sea World) within six months. However, the landmarks such as SDSU (the 100-unit contour polygons) and Sea World (the 10-unit contour polygons) (shown in Figure 3b) experience larger variations than established cities (shown in Figure 3a). Sea World has a wider activity boundary in summer since summer is the peak season when a great number of tourists visit it, including students. SDSU has a larger boundary in September and October compared to August. During the Fall semester, more active students have classes on campus. During the summer, many students go on vacation or internships. Few people and activities happen at this time, so few people can potentially send tweets. Unlike established cities, SDSU and Sea World are the two most representative places which show large spatial and temporal dynamics. Consequently, we selected a wider time frame to analyze the spatial-temporal change in SDSU and Sea World. More tweets data has been collected from January 1, 2015, until December 31, 2016, as well as monthly Sina Weibo data from October 1, 2013, until September 31, 2014. For the case study of Beijing, Peking University, Tiananmen Square, and Wangfujing have been selected for the analysis. These places are considered as highly dynamic places rather than established districts. A calendar year has been divided into four seasons: spring (March 1 to May 31), summer (June 1 to August 31), fall (September 1 to November 30), and winter (December 1 to February 28th/29th). Also, the search radius of KDE must be set uniformly for all four seasons. The search radius using for all seasons is the default search radius of the spring of each place name. Then, we also analyzed the seasonal KDE change with KDE normalization to refrain the effects of uneven data volume among different seasons.

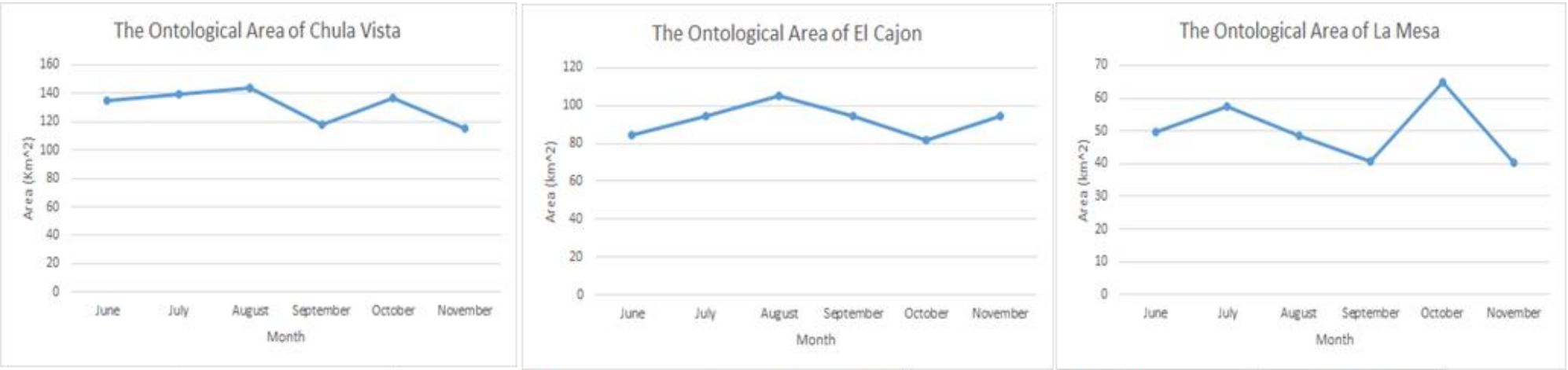

| Month | Area (km^2) |
|---|---|
| June | 135.228509 |
| July | 139.675634 |
| August | 143.624261 |
| September | 118.327127 |
| October | 136.686467 |
| November | 115.112074 |

| Month | Area (km^2) |
|---|---|
| June | 84.485433 |
| July | 94.629557 |
| August | 105.267927 |
| September | 94.101866 |
| October | 81.417959 |
| November | 94.150175 |

| Month | Area (km^2) |
|---|---|
| June | 49.580693 |
| July | 57.546231 |
| August | 48.515091 |
| September | 40.831049 |
| October | 64.844252 |
| November | 40.198377 |

***Figure 3a***. *The monthly changes of the ontological areas of Chula Vista, El Cajon, and La Mesa*

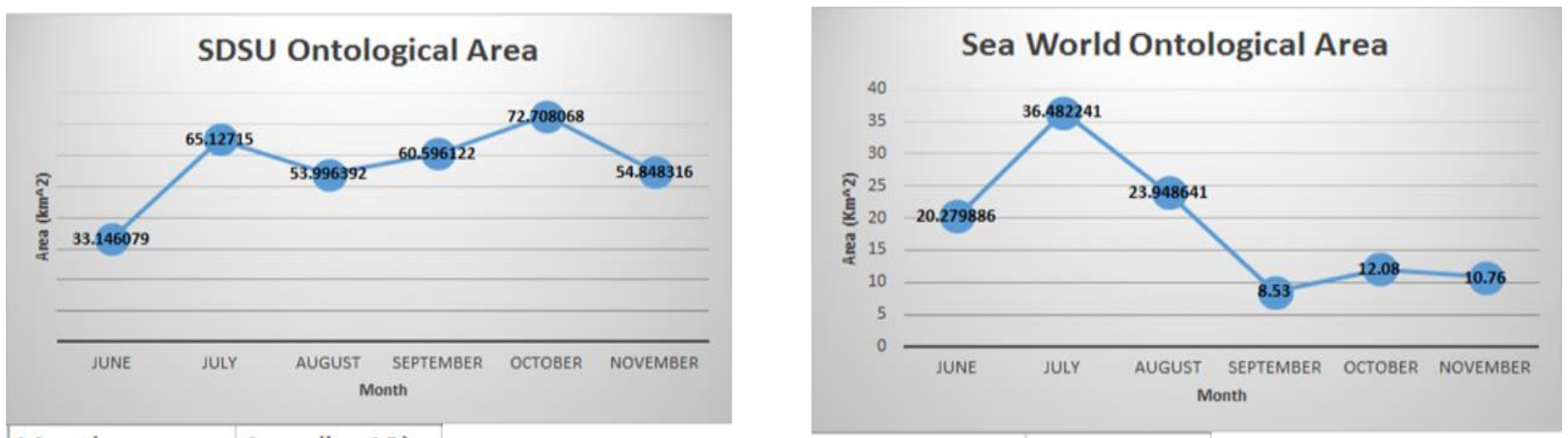

| Month | Area (km^2) |
|---|---|
| June | 33.146079 |
| July | 65.12715 |
| August | 53.996392 |
| September | 60.596122 |
| October | 72.708068 |
| November | 54.848316 |

| Month | Area (km^2) |
|---|---|
| June | 20.279886 |
| July | 36.482241 |
| August | 23.948641 |
| September | 8.53 |
| October | 12.08 |
| November | 10.76 |

***Figure 3b***. *The monthly changes of the ontological area of SDSU and Sea World*

## 3.4 Semantic Analysis

Semantic analysis is a method that use computer to process, analyze, then understand and interpret text on Internet, which is crucial for modeling place name ontology. Pointwise Mutual Information (PMI) method was applied to rank strongly related words in tweets collected using each original place name (Church and Hank 1990). The mutual information method considers giving the candidate words' association strength with the place name's concepts, and top words were selected as seed keywords. However, word frequency may not be able to indicate the significance of the word. Primitively, this analysis ran queries in MongoDB and SQL database to select all associated tweets within the same place name collection. The wordcloud stands for a high-frequency word list that records the top 50 words that have existed in the tweet text that associated with the place name. Second, we used the PMI algorithm to score the 50 highest frequency words (after removing stop words) for their precision in returning tweets connected with the place. Each place is accompanied by a list of seed keywords. The place name's ontology was generated based on two approaches: 1) Query the social media data from the place name itself only with the combination of both lists of associated seed keywords and the place name itself; 2) Analyze the semantic meaning of the place name's attributes. We applied word segmentation to calculate the PMI scores for Chinese Weibo and Tweets posts and generate wordcould.

$$PMI(x;y) \equiv log\frac{p(x,y)}{p(x)p(y)} = log\frac{p(x|y)}{p(x)} = log\frac{p(y|x)}{p(y)} \qquad (2)$$

*In formula 2, 'x' stands for the queried place name, and y stands for a high-frequency keyword associated with the place name. PMI is calculated by the logarithm of the probability of Weibo posts which contain both the queried place name the keyword p(x,y) over the production of the probability of*

*Weibo posts which contain the place name p(x) and the probability of Weibo posts which contain the keyword p(y).*

We focus on the wordcloud and PMI outcome's comparison between the inside of the region's place boundary circle and outside of the region's place boundary circle. The boundary circle type should be determined based on how well it fits within administrative boundaries and how easily and automatically it is generated. Both wordcloud and PMI calculations for all nine place in San Diego and eight place in Beijing have been used. Then, we explored semantic differences between the inside and the outside of the region. The flow chart of Semantic Analysis is as shown in the Figure 4.

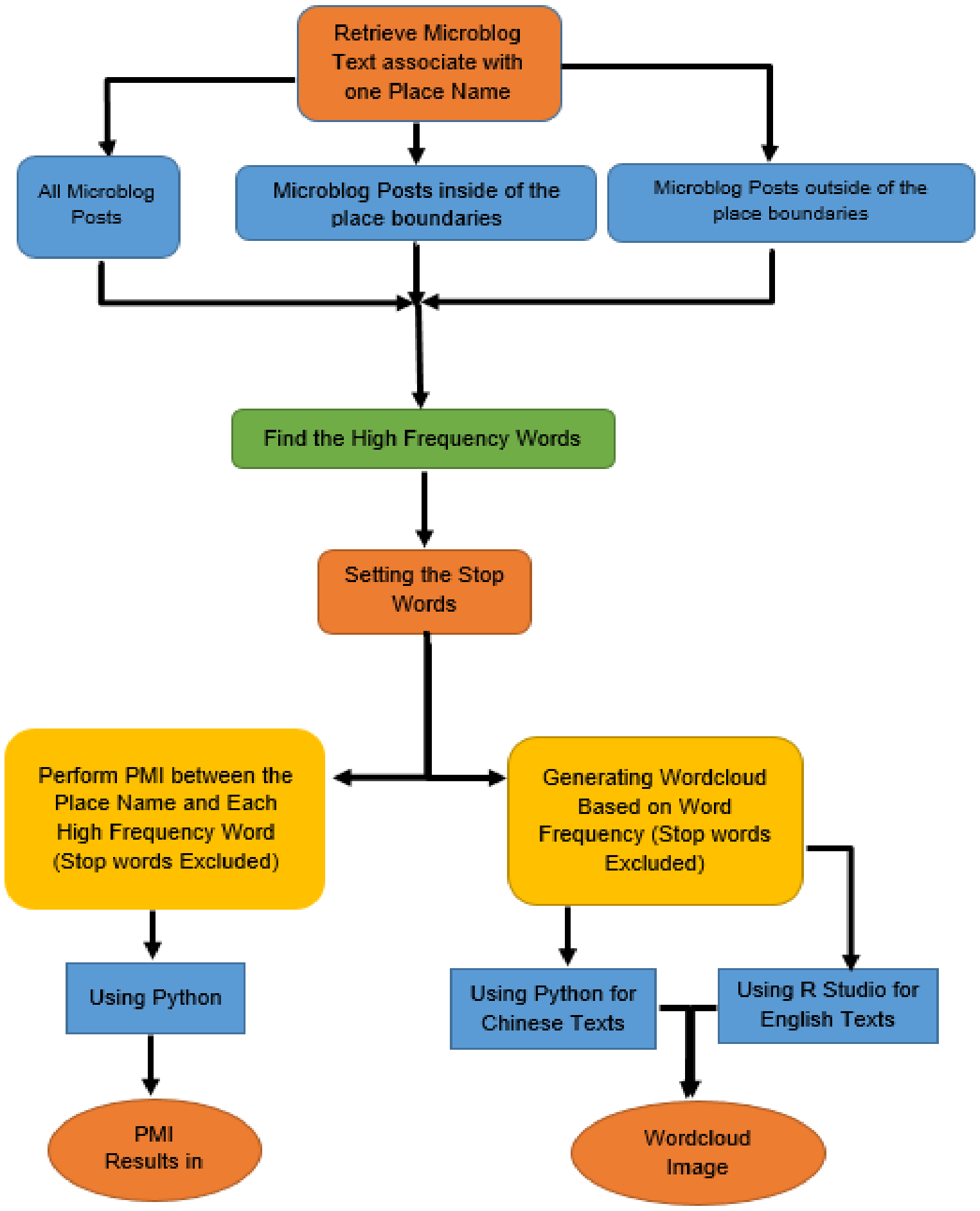

***Figure 4**. The flow chart of Semantic Analysis*

# 4. Results

## 4.1 KDE Analysis

The purpose of KDE normalization is to construct a solid and significant boundary in San Diego and Beijing. It automatically picks the 0 contour line as the boundary. Thus, it does not need to select one particular contour line as the boundary specifically. If the tweet density of specific regions is relatively higher than the density of total tweets, the region is more significant and represents the place name, and it is more likely to be inside of the boundary. The tool "raster algebra" has been used for KDE normalization and Formula 2 has also been applied on ArcGIS.

***San Diego***

In general, the total areas inside most places' normalized boundaries are smaller than the area inside of the settled un-normalized place boundaries. The boundaries are composed of multiple smaller circles. They appear as light blue on the following maps. The boundaries generated by normalized KDE analysis are called circled regions. The boundary does not contain the regions that have low Weibo post populations with a specific keyword.

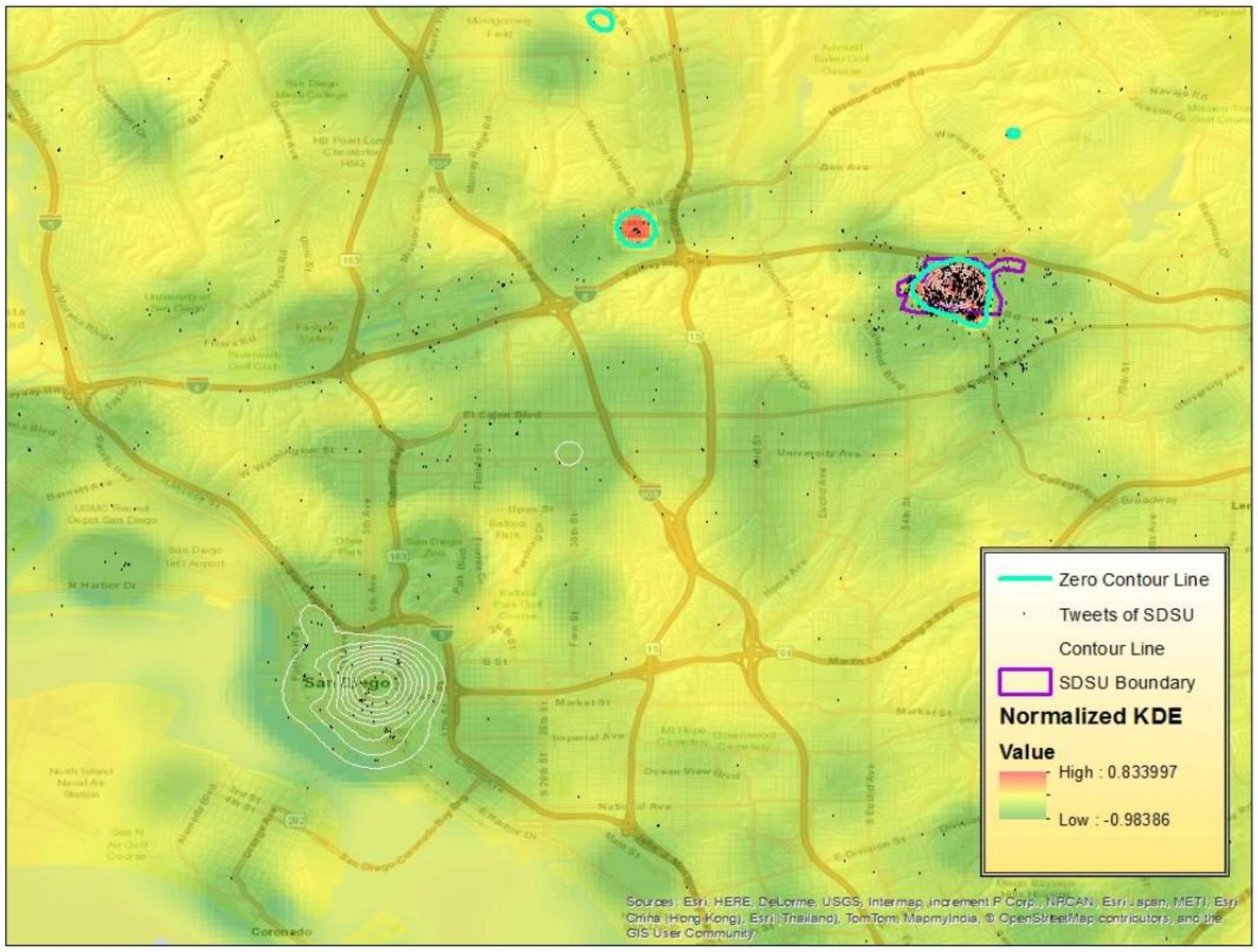

***Figure 5***. *Normalized KDE Output Result of SDSU*

Based on Figure 5, two major hot spots exist. The first one is located on the main campus of SDSU. The second one is at Qualcomm stadium. The boundary at the main campus of SDSU fits with the administrative boundary well. Unlike the un-normalized contour line boundary, the normalized KDE contour line boundary eliminated the Downtown area. The result suggests that certain kinds of places might have disjoint "activity zones" that are part of the place concept, in the sense that they are part of ordinary human activity connected with the place, but actually they are spatially disjoint. There is a coverage difference between the output result of a normalized boundary and an un-normalized boundary of San Diego. The un-normalized boundary (contour line 10) does cover the boundary of the city of San Diego. However, the normalized boundary almost covers the whole county area.

***Beijing***

In Figure 6, the main boundary circle is located northeast of Tiananmen Square. There is a small portion located inside Tiananmen Square and is very close to Tiananmen. People might be more likely to post relevant weibos around the Eastern Tiananmen subway station rather than Tiananmen Square itself. The boundary of district-level place is composited by a large number of smaller circles that are located away from the densely populated zones. By contrast, none of the circled regions are located inside the administrative boundary of Peking University. For Tiananmen Square and Wangfujing, the real administrative boundary is located west of their corresponding circled region.

Overall, the KDE method with normalization executes boundary portrayal more intelligently than without normalization. Most of the normalized boundary circles fit into the administrative boundaries. However, the output results are not ideal for high-level places in place hierarchy. It is still more meaningful to apply the normalized boundaries to perform regional semantic analysis (PMI and wordcloud) in the next step.

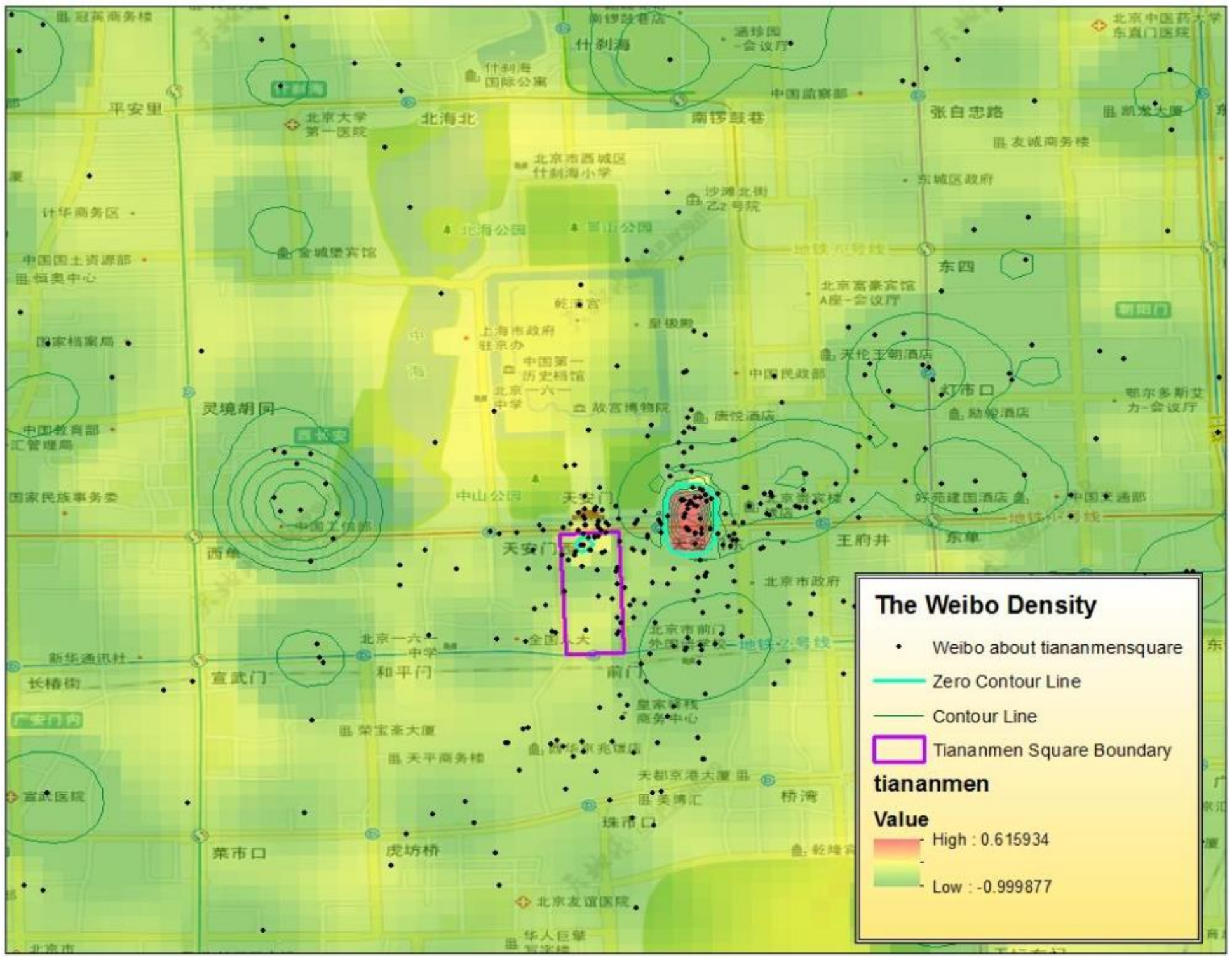

***Figure 6.*** *Normalized KDE Output Result of Tiananmen Square*

# 4.2 Hierarchal Clustering and DMDBSCAN.

Five places in San Diego (Chula Vista, El Cajon, La Mesa, SDSU, and Sea World) and five places in Beijing (Chaoyang, Haidian, Peking University, Tiananmen Square, and Wangfujing) are chosen to conduct the clustering analyses and some of the cluster results are shown below.

***San Diego***

This algorithm has only been applied to five places in San Diego. Concave and convex hulls are generated based on the largest existing cluster. According to the following output maps (Figure 7a-b), the established cities have much smaller convex and concave hulls than their administrative boundary in general. However, the results are opposite at other types of places such as SDSU and SeaWorld. They have larger convex and concave hulls than their actual boundaries.

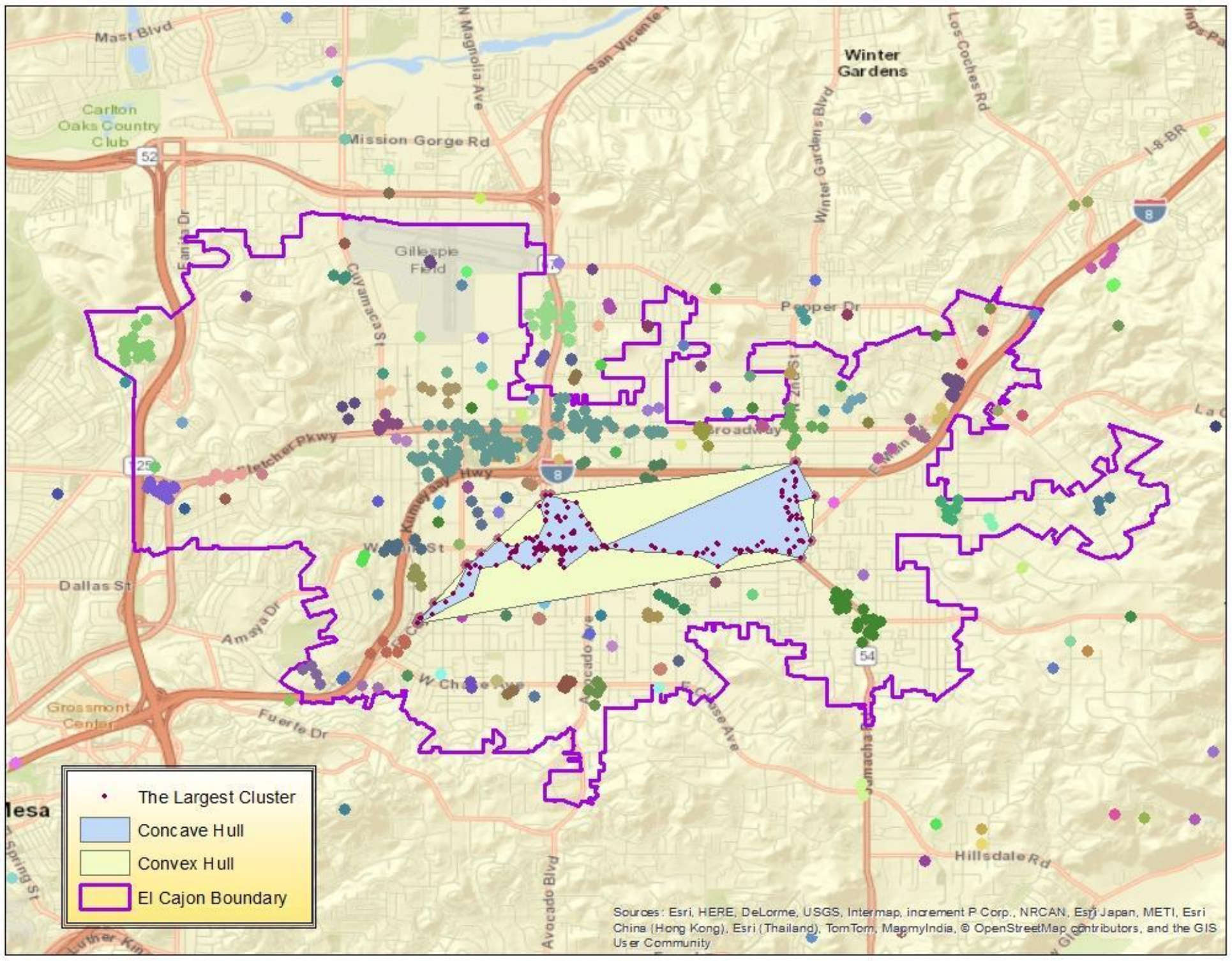

*Figure 7a*. *Hierarchal Clustering Output Result of El Cajon*

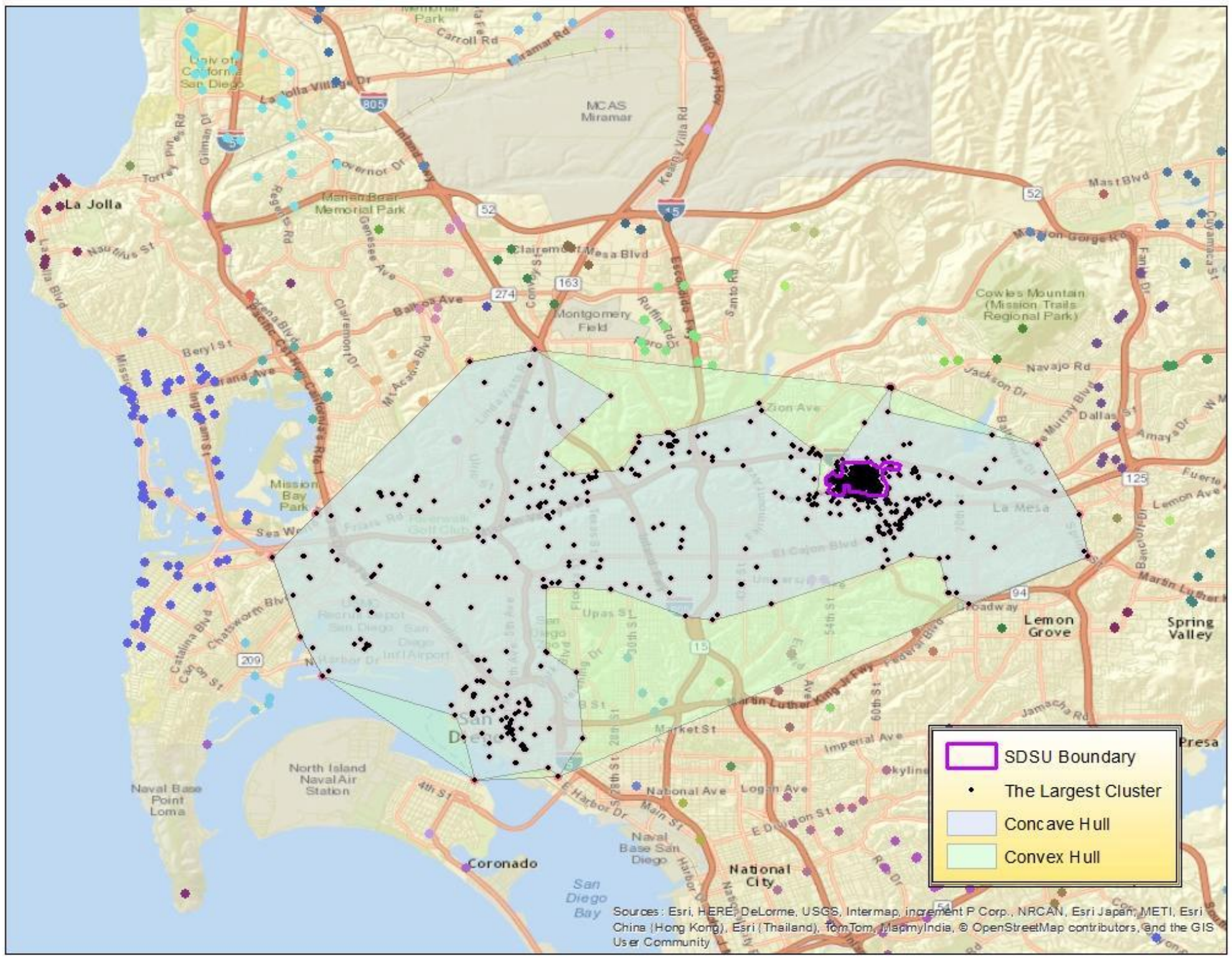

***Figure 7b***. *Hierarchal Clustering Output Result of SDSU*

### ***Beijing***

Hierarchal clustering has been applied to four place in Beijing. The following two maps (Figure 7 c-d) demonstrate the boundaries of two places (Haidian and Wangfujing). Only four places show the hierarchal clustering output results. However, the boundaries of Wangfujing and Tian'anmen square are not ideally fit the administrative boundaries. Similar to the case study of San Diego, there are no remarkable areal and size differences between the concave hull and convex hull output.

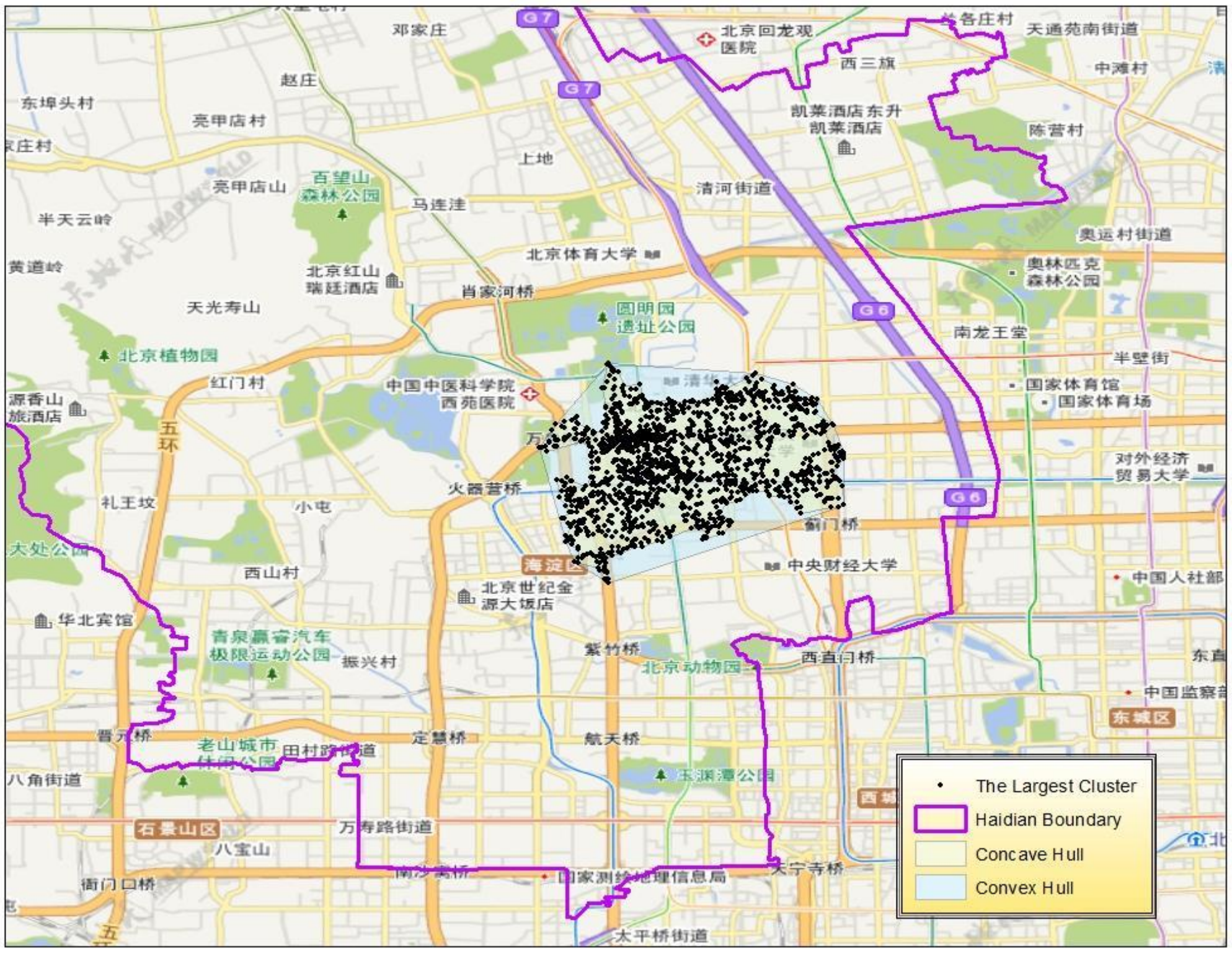

***Figure 7c***. *Hierarchal Clustering Output Result of Haidian District*

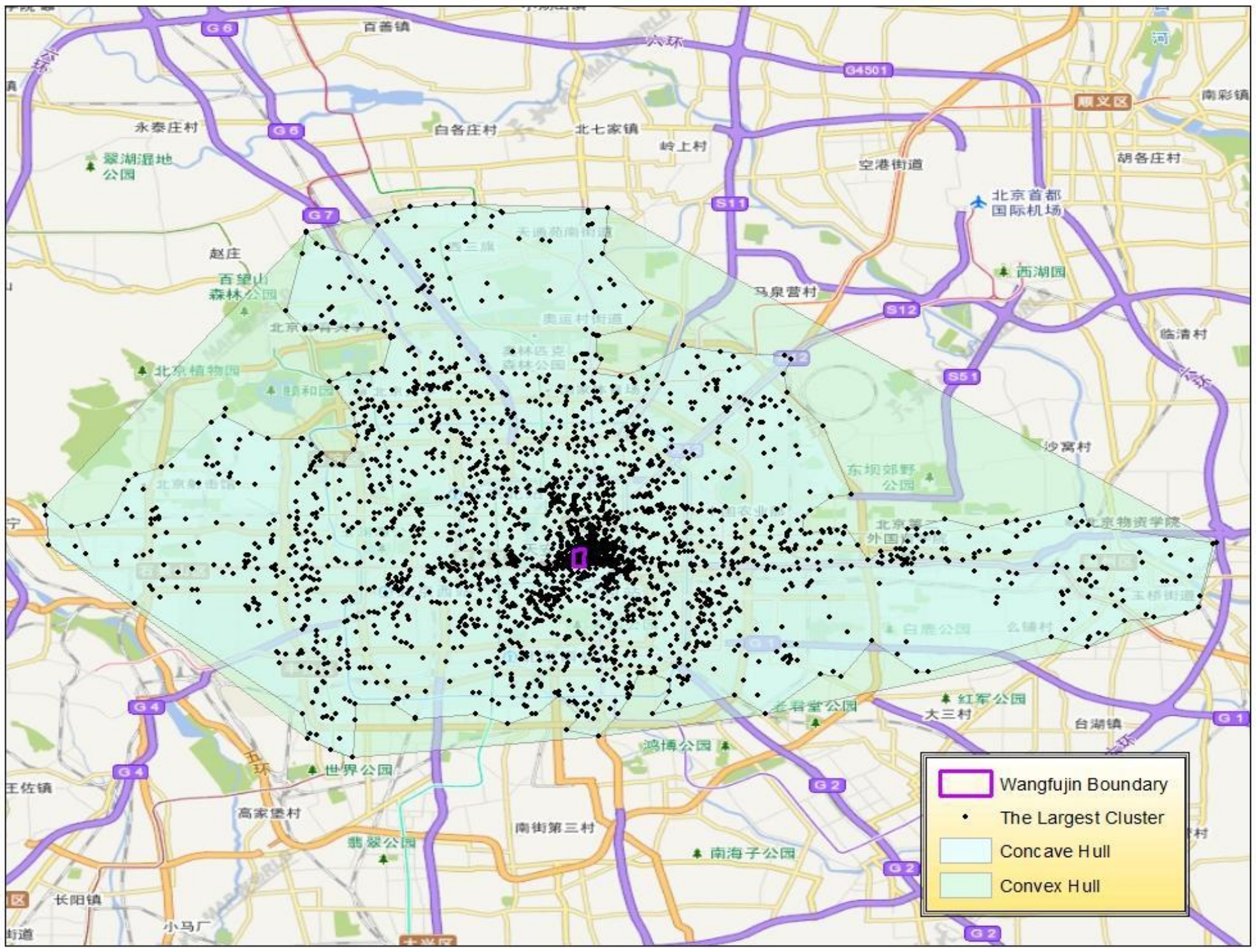

***Figure 7d***. *Hierarchal Clustering Output Result of Wangfujing*

Overall, the output of hierarchal clustering does not ideally portray the appropriate boundaries. In the case study of San Diego, the boundaries of established cities in San Diego are mostly much smaller than actual boundaries (Figure 12a). However, non-city places such as SDSU have relatively large boundaries. In the case study of Beijing, the non-city places such as Wangfujing have extraordinarily large boundaries (Figure 12d). Even though the results are not satisfactory, their locations are precise if the boundaries are smaller than the administrative boundaries.

As Figure 8 portrays, the purple line indicates the administrative boundary of  Chula Vista. Points from multiple clusters may be close to each other. There is no obvious difference between the two largest clusters. Concave and convex hulls can not be generated based on the largest existing cluster using DMDBSCAN.

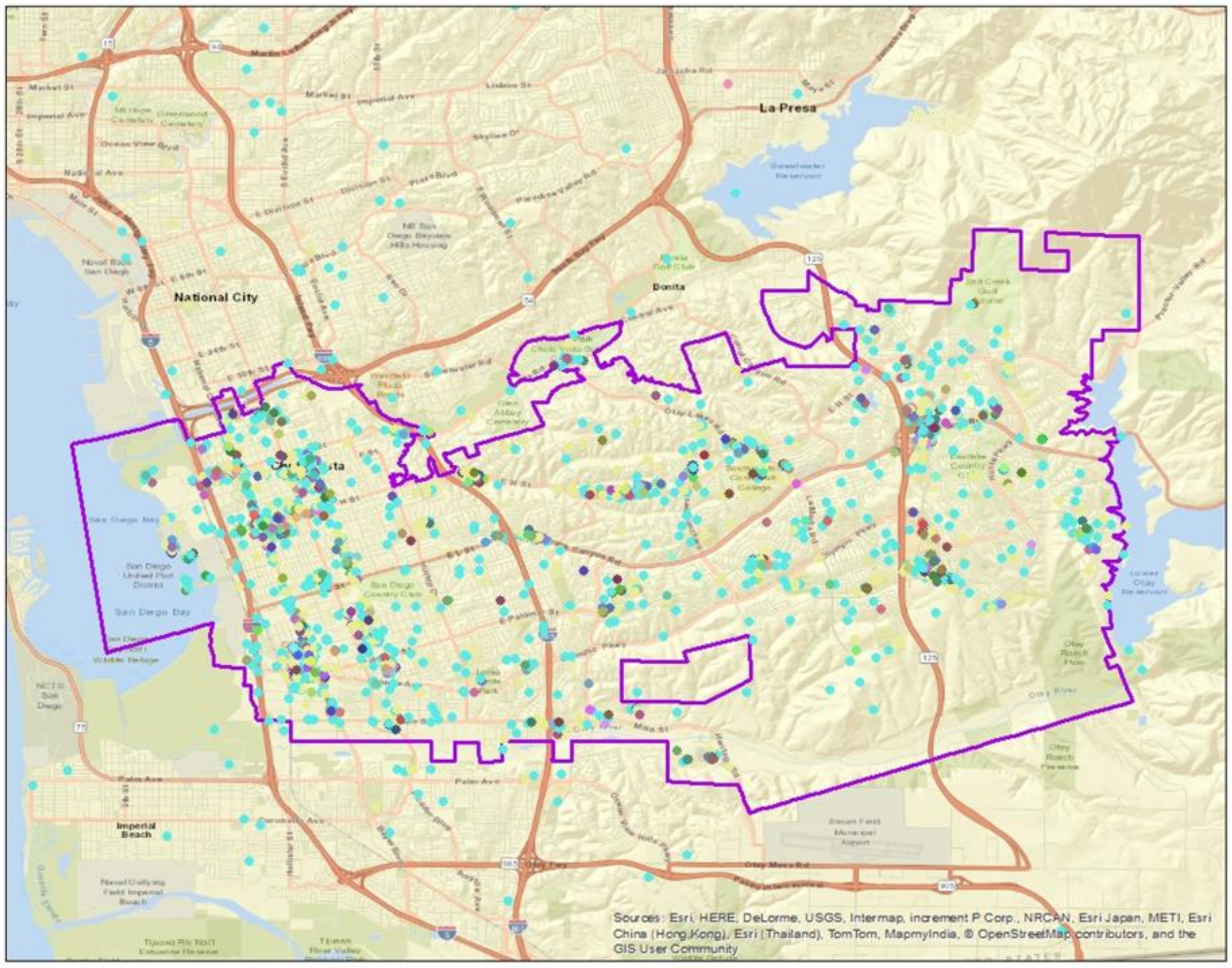

***Figure 8***. *DMDBSCAN Clustering Result of Chula Vista*

## 4.3 Spatial-Temporal Analysis

The non-established cities showed different temporal patterns than established cities. For instance, the boundary showed a dramatic change between Spring and Summer. During the summer break at San Diego State University, the size of the ontological boundary is smaller than it in Spring or Fall semester months. There are fewer students on campus in the summer, so they tweet less frequently. On the other hand, the extent of geo-tagged tweets is also small because some campus services are shut down and fewer events are held during the summer. Therefore, people cannot tweet there.

### ***Quarterly Changes***

The quarterly change of absolute can be retrieved by subtracting the previous season's Weibos' density value from the later season's Weibos' density value. And we normalized the absolute change by population. The following maps (Figure 10 and 11) showed the normalized KDE raster layers. The red pixels represent those regions that have relative Weibo posts increase, and the green pixels represent regions with a decrease.

#### ***San Diego***

The following four maps (Figure 10) show the seasonal changes at San Diego State University. the normalized KDE values change dynamically on campus, Qualcomm stadium, and the downtown area. From spring to summer (Figure 10a), the KDE values inside and around the campus area and downtown area have a dramatic decrease. Most SDSU students may not be present during the summertime. They might visit their families or find a part-time job or internship. From Summer to Fall (Figure 10b), the KDE value increases significantly in the campus area and Qualcomm stadium. From Fall to Winter (Figure 10c), the central campus region and Qualcomm stadium show a decrease, but the surrounding area shows an increase.

From winter to spring (Figure 10d), the SDSU campus and the northwest part of downtown show dramatic increases, but the Qualcomm stadium shows a decrease.

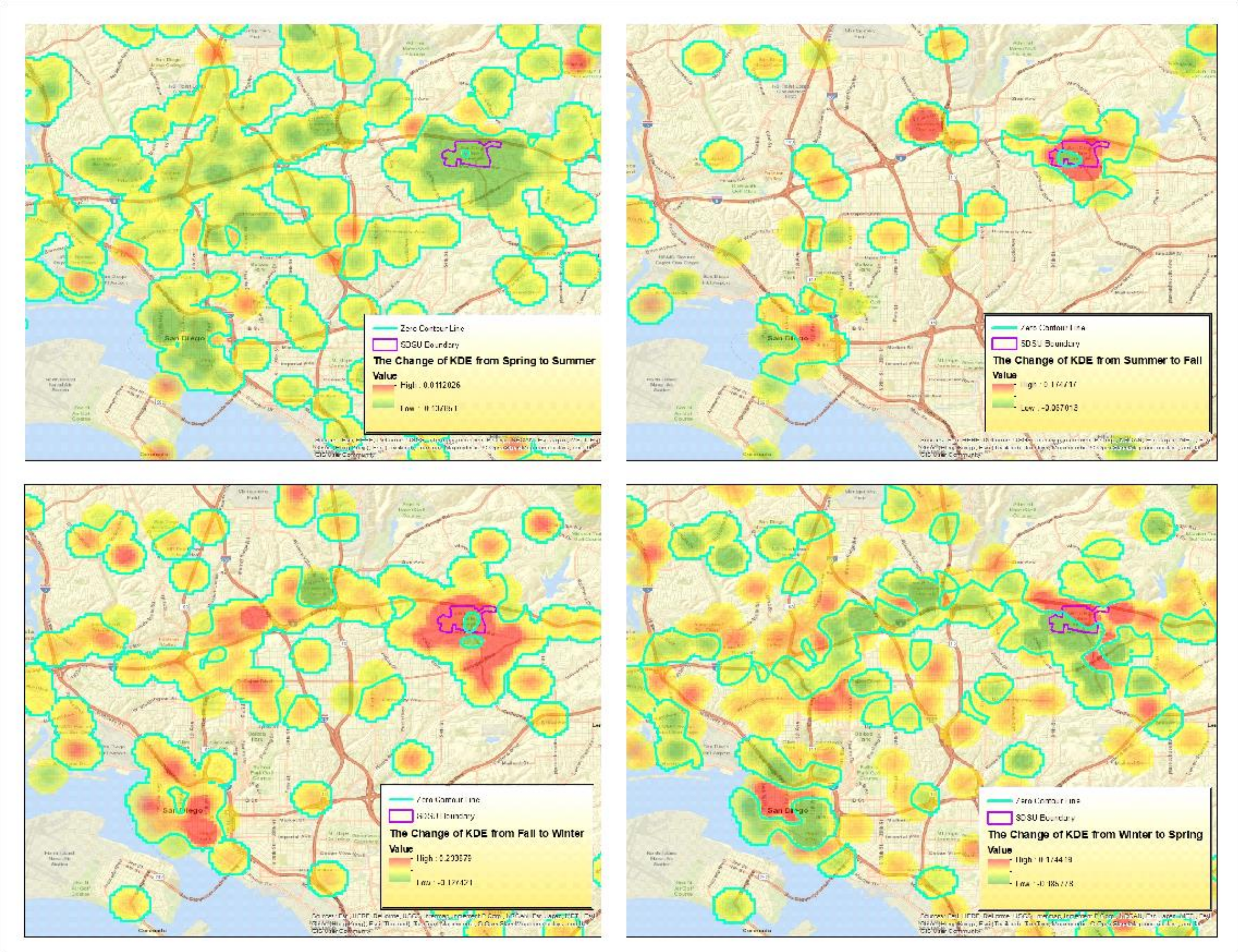

*Figure 10. Seasonal Changes of SDSU using normalized KDE value. a). spring to summer; b). summer to fall; c). fall to winter; d). winter to spring.*

### *Beijing*

The following four maps (Figure 11 a-d) indicate the absolute KDE value changes of all Weibo tweets about Peking University. From Spring to Summer, there is a significant decrease on campus and the surrounding area east of the campus. Students tend to post fewer Weibo posts during the summer. The reason might be that students need to visit their families and find internships during the summer break. Also, there is a certain amount of students who graduate and do not register courses in the summer. This phenomenon is similar to the case of SDSU. An increasing trend appears from summer to fall, but the decrease also exists in the north part of the campus. From winter to spring, the largest increase is shown on the map. Students usually come back for school in early September and late February.

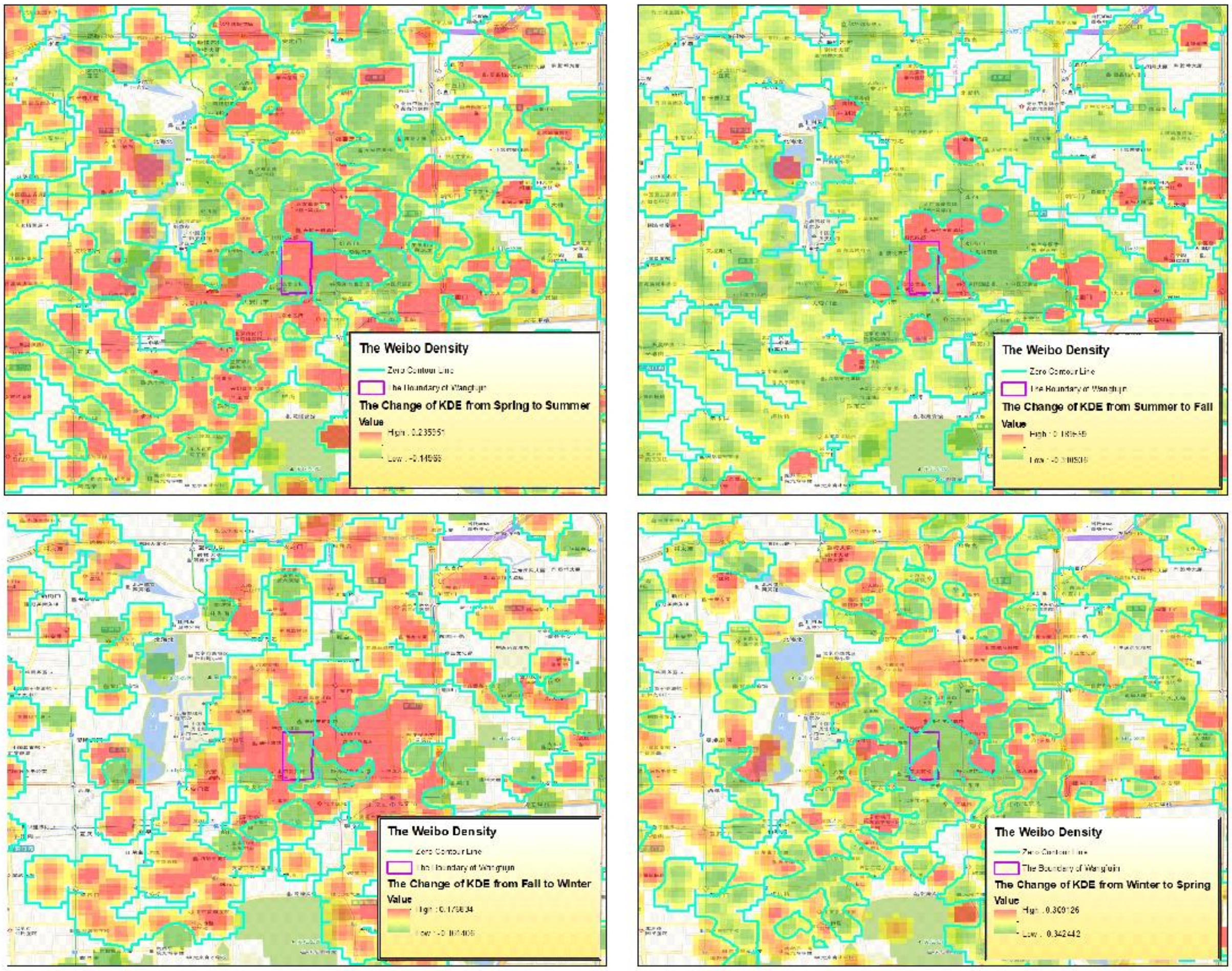

*Figure 11. Seasonal Changes of Peking University using normalized KDE value. a). spring to summer; b). summer to fall; c). fall to winter; d). winter to spring.*

## 4.3 Semantic Analysis

***San Diego***

Table 4 shows the PMI result in Chula Vista. Overall, the full wordcloud (Figure 12) looks similar to the in-circle wordcloud. The out-circle wordcloud shows some remote place. The amphitheater is one of the most high-frequency words in both the full and in-circle wordclouds. The amphitheater is located in the city of Chula Vista. Lots of large concerts are hosted there. East Lake is another community in Chula Vista. Two Costco wholesale markets are located in Chula Vista. "Otay" Ranch is obviously shown in the out-circle wordcloud since Otay Ranch is a large community in Chula Vista.

The PMI results show similar patterns to the three wordclouds. However, there are several insignificant words that appear in the out-circle PMI result list but cannot be found in the out-circle wordclouds. For example, the word "Pius" has a very high PMI score of 5.397617, but it has low word frequency. The wordclouds only highlights high-frequency words. Therefore, the word "Pius" may not be able to be shown in the out-circle wordclouds.

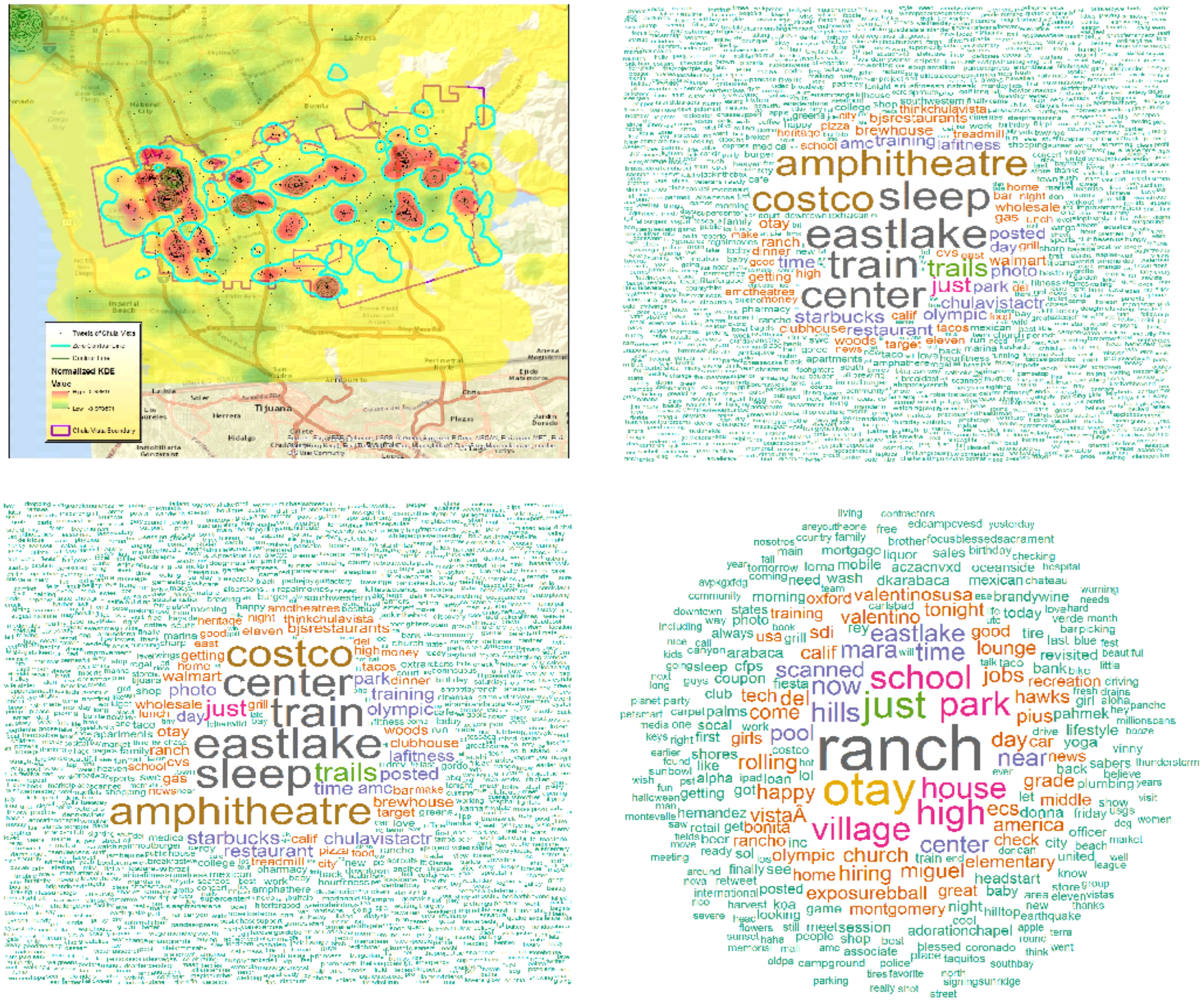

Figure 12. a). Normalized KDE Output Result of Chula Vista; b). Full wordcloud of Chula Vista; c). In-circle wordcloud of Chula Vista; d). Out-circle wordcloud of Chula Vista.

*Table 4. PMI Result of Chula Vista*

| High Frequency Words | Full PMI Score | Word Frequency | High Frequency Words | In-Circle PMI | Word Frequency | High Frequency Words | Out-Circle PMI | Word Frequency |
|---|---|---|---|---|---|---|---|---|
| Eastlake | 5.262274 | 2238 | @chulavistactr | 6.233865 | 261 | Pius | 5.397617 | 30 |
| Costco | 4.741455 | 3165 | @Costco | 5.502333 | 478 | Otay | 3.830088 | 2899 |
| Amphitheatre | 4.573374 | 4252 | Eastlake | 5.367245 | 2160 | Village, | 3.625219 | 421 |
| Trails | 4.274415 | 3140 | Wholesale | 5.315304 | 441 | @exposurebball | 3.482329 | 188 |
| Sleep | 4.184739 | 7114 | Clubhouse | 5.177324 | 512 | ECS | 3.437114 | 377 |
| Train | 3.830405 | 12366 | @bjsrestaurants | 4.885459 | 751 | Rolling | 2.94733 | 1284 |
| Otay | 3.782087 | 3188 | @LAFitness | 4.841126 | 934 | CA) | 2.692048 | 112601 |
| 10mi, | 3.697726 | 6281 | Costco | 4.820489 | 3144 | CA | 2.545377 | 369655 |
| cloudy, | 3.60091 | 7849 | Brewhouse | 4.597081 | 1038 | Center | 2.508469 | 33522 |
| humidity, | 3.535073 | 13857 | Amphitheatre | 4.592416 | 4063 | scanned | 2.432153 | 2821 |

### *Beijing*

Based on the three wordclouds (Figure 13), they are distinct from each other. Some terms are highly associated with Tiananmen Square such as "Forbidden City" (故宫), "raising the flag" (升旗), and "security check" (安检). Raising the flag is the routine activity that happens every morning when the sun comes out. Many residents and tourists are willing to watch the raising when they come to Tiananmen Square. A security check is necessary for entering Tiananmen Square. Forbidden City is behind Tiananmen or north of Tiananmen Square.

The three PMI tables also show similar patterns to the wordclouds.  The words "raising the flag" (升旗), "lowering the flag" (降旗), "raising the national flag" (升国旗), and "national flag" (国旗) are the top 4 keywords in the in-circle PMI table. There are many keywords associated with tourism that are highly ranked in the out-circle PMI table such as "travel agency" (旅行社), "salesman" (业务员), "tourist" (游客). The semantic meaning changes significantly through spaces about Tiananmen Square.

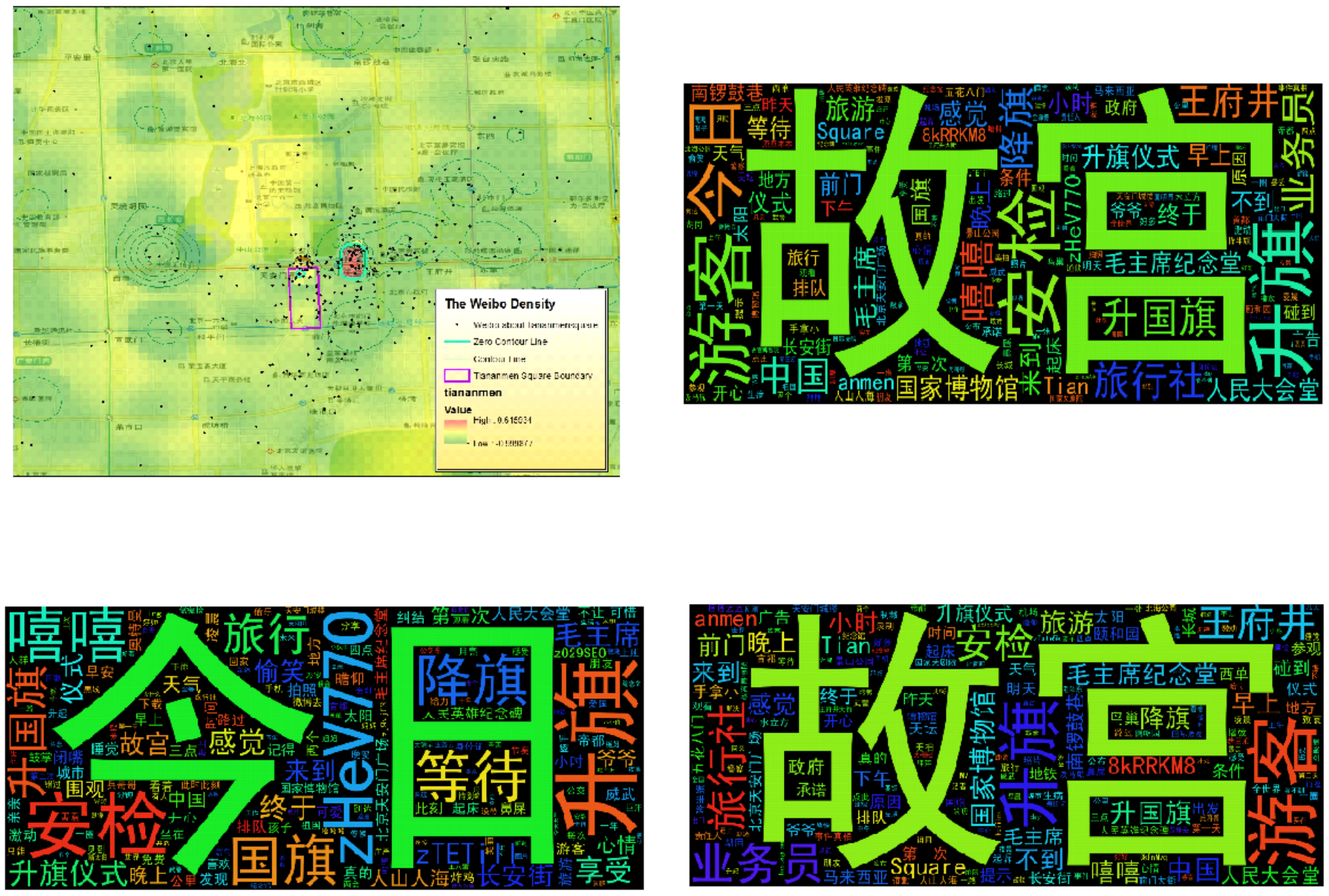

Figure 13. a). Normalized KDE Output Result of Tiananmen Square; b). Full wordcloud of Tiananmen Square; c). The wordcloud of Tiananmen Square (Inside the zero value circle); d). The wordcloud of Tiananmen Square (Outside the zero value circle).

*Table 5. PMI Result of Tiananmen Square (天安门广场)*

| High Frequency Words | PMI Score | Word Frequency | High Frequency Words | In-Circle PMI Score | Word Frequency | High Frequency Words | Out-Circle PMI Score | Word Frequency |
|---|---|---|---|---|---|---|---|---|
| 升旗 | 5.58007992 | 3916 | 降旗 | 5.987234 | 1016 | 业务员 | 6.274158 | 373 |
| 游客 | 4.88094603 | 2337 | 升旗 | 5.58008 | 3916 | 旅行社 | 5.648951 | 697 |
| 故宫 | 4.73611517 | 10226 | 升国旗 | 5.491129 | 1723 | 升旗 | 5.58008 | 3916 |
| 安检 | 4.20768799 | 6022 | 国旗 | 5.101448 | 4124 | 游客 | 4.880946 | 2337 |
| 他们 | 1.17128614 | 89989 | 安检 | 4.207688 | 6022 | 故宫 | 4.736115 | 10226 |
| 看 | 0.78511073 | 1085317 | 今日 | 2.523698 | 33851 | 北京 | 1.268972 | 634364 |
| 今天 | 0.68423151 | 521481 | 等待 | 1.698638 | 41843 | 他们 | 1.171286 | 89989 |
| 走 | 0.681283 | 387258 | 看 | 0.785111 | 1085317 | 看 | 0.785111 | 1085317 |
| 去 | 0.62707674 | 998577 | 今天 | 0.684232 | 521481 | 今天 | 0.684232 | 521481 |
| 到 | 0.55364866 | 1456480 | 去 | 0.627077 | 998577 | 走 | 0.681283 | 387258 |

All of the full wordclouds reflect the properties, place category, or most remarkable landmarks associated with the place in the case study of San Diego. Also, most of the full wordclouds describe many alternative terms for a place. The wordcloud output and PMI results have many similarities between these two place. Even though a certain amount of residential, communities, universities, and business centers are located in these two districts, the wordclouds and PMI results do not properly reflect these attributes. The semantic analysis of the places indicates the hierarchal relationship of place, which can be used for studying the place name hierarchy.

Overall, in the case study of San Diego, the in-circle and out-circle wordclouds and PMI lists are distinct from each other, but the in-circle wordcloud and PMI results are similar to the full wordcloud and PMI results. However, in the case study of Beijing, the other results are not consistent with the full wordclouds and PMI results. Thus, the semantic meanings of place vary spatially in our case studies.

# Discussion

Kernel Density Estimation (KDE) analysis, hierarchal clustering, and line feature processing can be used to define place name boundaries or footprints with geo-tagged social media datasets. The normalized KDE analyses generate the boundaries of place in different ways. The KDE search radius can be used to identify the feature type (Point, Polyline, or Polygon). Most boundaries and footprints created by the KDE algorithms are closely attached to the true locations or official boundaries of place.

Different types of place illustrated different spatial-temporal patterns of their ontology models. Two examples (S.D.S. U and Sea World) on Twitter and three examples (Peking University, Tiananmen Square, Wangfujing) on Sina Weibo are selected to demonstrate the dynamic change of spatiotemporal patterns of place ontology. We utilized the normalized KDE method to trace the post values' changes between two consecutive seasons. In general, the seasonal KDE changes give an intuitive demonstration that reflects human activities of a place. For example, the seasonal KDE changes of SDSU and Peking University show major student activities throughout different seasons. Significant decreases in KDE are shown on the main campus of SDSU and Peking University from spring to summer due to summer vacation, and the maps also portray large increases from summer to fall and from winter to spring because students are back to school. The seasonal KDE change of Sea World shows the dynamic of tourists' activities. A major significant increase appears from spring to summer and a decrease appears from summer to fall because summer is the peak season for recreational travels

Based on the output result of PMI calculation and wordcloud, different place have different semantic meanings. Wordcloud and PMI calculation can briefly list multiple alternative terms about that specific place. For example, many places have their nicknames and famous landmarks inside the wordcloud. These terms are possibly listed on the PMI list or exaggeratedly appear on the wordcloud. However, the semantic meaning can change throughout different locations (within the boundary of place or outside the boundary of place). By viewing the place hierarchy, higher-level places' names cover some lower-level places' names.

In summary, this research contributed new ideas about how to use G.I.S. analysis and social media data to spatially and semantically represent a place ontology. A new algorithm was introduced to identify the feature type and the spatial footprint of place by using geo-tagged social media. The parallel analysis of normalized KDE analysis indicated the dynamic characteristics of place name ontology over space and time. We introduced and utilized two effective methods of semantic analysis (PMI and wordcloud) that produce meaningful attribute information associated with place. Since this research is interdisciplinary, future research will further explore the application of social media on ontology and intersect with advanced knowledge from linguistics and computational social science.

# Conclusion

In this research, two popular and highly dynamic microblog social media platforms, Sina Weibo and Twitter, were utilized to build crowd-sourced place ontology in Beijing and San Diego. Since many people like to share their activities and messages on social media platforms, researchers can collect social media data and perform data mining and linguistic analysis to monitor place-related human behaviors spatially and semantically. Specifically, compared to the unit boundaries in Google maps, social media is faster at detecting ontological changes: for example, in Google maps, SDSU refers to SDSU's main campus; when social media data is used to extract tweets with SDSU keywords and extract ontological boundaries, researchers find that new SDSU stadiums far from the main campus are also detected. Dynamic ontological boundaries can be detected by crowd-sourced geo-tagged social media data, which can illustrate the dynamic change of place from a spatiotemporal perspective.

This research combines social media, cartography, semantic analysis, and spatial-temporal analysis to build place name ontology. The major contribution of this research is to analyze the inter-relationships between place, space, and their attributes in the field of geography. Researchers can use crowd-sourced data (social media) to study the ontology of places rather than relying on traditional gazetteers. By testing with different algorithms for spatial analysis, some methods can generate better ontological boundaries of specific places. Additionally, this study illustrates the dynamic changes in place ontology due to the change in human activities and conversation over time and space.

This research utilized Twitter data and Weibo datasets to analyze the trends and patterns of each place name spatially, temporally, and semantically. Before applying the social media data, the procedure of data cleaning is required to remove noises and advertisements. Our analysis outcome suggests the place name ontology is dynamic and variant. Different algorithms will generate distinct boundary results. This research also demonstrates that the boundary of places is dynamic over time. This study focused on the semantic and spatiotemporal analyses of dynamic ontological models using various types of place. We expect this model to be applied to urban planning in the future.